**PERSPECTIVES ON A SOLID STATE NMR QUANTUM COMPUTER**

Edward B. Fel'dman[1,2] and Serge Lacelle[2*]

[1] Institute of Problems of Chemical Physics, Russian Academy of Sciences, 142432 Chernogolovka, Moscow Region, Russia.

[2] Département de Chimie, Université de Sherbrooke, Sherbooke, Québec, Canada J1K 2R1

[*] To whom correspondence should be addressed.

**ABSTRACT**

A quantum information processing device, based on bulk solid state NMR of the quasi-one dimensional material hydroxyapatite, is proposed following the magnetic resonance force microscopy work of Yamamoto et al (quant-ph/0009122). In a macroscopic sample of hydroxyapatite, our solid state NMR model yields a limit of $10^8$ qubits imposed by physics, while development of current technological considerations should allow an upper bound in the range of hundreds to thousands of qubits.

The physical implementation of quantum information processing (QIP) devices has attracted a certain number of interesting proposals [1]. At present, ensemble quantum computing with liquid state Nuclear Magnetic Resonance (NMR) offers the largest quantum register in a 7-qubit molecule, the design of which was suggested by one of us (S.L.) [2]. The recent magnetic resonance force microscopy work of Yamamoto et al [3-6] has inspired us to ask "Could we design a solid state NMR QIP device with the quasi-one-dimensional material hydroxyapatite or one of its modified derivatives [7] ?" Our efforts to answer this question, as presented here, have been strongly influenced by NMR work on this material from the 1960s up to today [ 6, 8-10]. While we realize that other solid state NMR QIP device proposals exist with modest numbers of qubits [3-6,11,12], the unique combination of circumstances in our design suggests that in a macroscopic sample of hydroxyapatite, the ultimate physical limit of $10^8$ qubits could be achievable in principle with bulk NMR methods. Physical estimates with current technology, as described below, indicate that hundreds to thousands of qubits should be within reach soon.

A crystalline sample of calcium hydroxyapatite, $Ca_5(PO_4)_3OH$ , 3.5 cm x 9.5 cm x 9.5 cm, contains approximately $10^{24}$ hydrogens. The microscopic structure of such a sample consists of one-dimensional (1-D) chains of hydrogens, from hydroxyl groups, with a lattice spacing of 3.44 Å, and with each chain surrounded by six nearest-neighbor (nn) parallel chains at a distance of 9.42 Å; calcium and phosphate ions are interspersed among the chains [7]. When the hydrogen chains are parallel to a strong Zeeman magnetic field along the z-direction, the sample has $10^8$ planes oriented perpendicularly to the field, with $10^{16}$ hydrogen nuclear spins in each plane. Thus, by adding a static magnetic field gradient along the z-direction, each plane of spins would be identified with a different resonant Zeeman frequency. While field gradients of 2 x 10$^4$ Gauss/cm are feasible [13], increasing such gradients by one hundredfold would produce a 300 Hz difference in the resonance frequencies of the $^1$H spins between each pairs of nn planes. Broadband decoupling of the dipole-dipole interactions across the sample with Lee-Goldburg averaging techniques [14], results in each plane consisting of an ensemble of uncoupled spins, the statistical state of which corresponds to a mixed state qubit. If the Lee-Goldburg line narrowing sequence irradiates uniformly a bandwidth of 45 kHz about the resonant frequency, then roughly 150 contiguous planes are decoupled uniformly; these individual planes would be the working qubits. Single qubit rotation is achievable with selective rf excitation of different planes. Two-qubit gates, arising from qubit-qubit interactions, are devised by selectively reintroducing dipolar interactions with rf irradiation, as outlined below, between nn, next-nearest neighbor (nnn) planes and beyond, up to a limit which depends on the value of the dipolar coupling used in the gate and possibly other competing dipolar interactions. Interactions between further qubits require a network of swapping operations as described in [15] and could possibly be implemented with NMR [16,17]. Work at low temperatures would reduce the relaxation rates in "clean" samples and enhance the NMR induction signals during the operations of this QIP device.

In our QIP model, two-qubit operations depend on dipole-dipole interactions among the spins on different planes. For a sample as just described, the strongest dipolar interactions are the intra-chain (or nn inter-plane) nn couplings with a value of about 3 kHz. Since the two-qubit logical operations will be of short duration, we consider the mean-field approximation where only the direct nn couplings between planes will be retained for the NMR spin dynamics, while others are neglected.

These requirements are satisfied by selecting the 375 Hz nn dipolar couplings among nnn planes; simple calculations show that nnn dipolar couplings among nnn planes have a value of 2 Hz, while nn couplings within a plane are 73 Hz. These latter couplings, and actually all intra-plane couplings, can be averaged out with MREV-8 multiple pulse decoupling [14] when applied selectively to a plane; note that the inter-plane dipolar couplings, being of "heteronuclear character" due to the presence of the magnetic field gradient, are not averaged out, but only scaled by this pulse sequence when applied selectively. In the presence of the Lee-Goldburg broadband (or non-selective) dipolar decoupling sequence, all dipolar interactions among the spins belonging to the 150 planes are averaged to zero in the sample described above. In order to reintroduce the dipolar couplings between spins on selective planes A and B, which are not necessarily adjacent, selective irradiations of the two planes are performed with the following procedure: The planes A and B are irradiated with each a rf field oriented at 144.7° with respect to the z-axis of the strong external field, with respective resonance frequencies $\omega_A$ and $\omega_B$. In a doubly rotating frame, with the $\bar{z}$-axis along the magic angle direction, the effective fields are along the perpendicular $\bar{x}$-direction. As a consequence, the secular part of the dipolar interactions of the spins of the planes A and B with spins belonging to other planes is averaged out, while the averaging of the dipolar interactions between the spins of the two selected and doubly irradiated planes A and B is incomplete. One can demonstrate the retention of dipole-dipole interactions of the form $D_{AB} I_{A\bar{x}} I_{B\bar{x}}$, where $D_{AB}$ is the nn dipolar coupling between spins belonging to planes A and B in the doubly rotating frame. The resulting Hamiltonian describing this situation is simply

$$\mathcal{H} = \hat{\omega}_A \sum_i I_{A\bar{x}i} + \hat{\omega}_B \sum_j I_{B\bar{x}j} + \sum_{i,j} D_{AB} I_{A\bar{x}i} I_{B\bar{x}j}$$

in complete analogy with the corresponding Hamiltonian of ensemble quantum computing in liquid state NMR [1]. The recent demonstration that the presence of any two-qubit interactions with accessible single qubit rotations ensures that universal quantum computation can be performed [18] with our proposed device.

Limitations exist on NMR ensemble quantum computing possibilities due to scaling considerations [19] and entanglement requirements [20,21]. By working at mK temperatures with the above QIP device, it should be possible to alleviate these problems to the extent that larger NMR signals will provide the opportunity to build quantum registers with very large numbers of qubits due to the architecture of this QIP device.

Several additional features of our proposed QIP device need to be addressed. The number of potential qubits is determined by an interplay between the finite excitation bandwidth of the Lee-Goldburg broadband decoupling technique and the strength of the magnetic field gradient which spreads the resonance frequencies. In the discussion above, the finite excitation bandwidth of the Lee-Goldburg broadband decoupling technique showed that only 150 contiguous planes were required, thus implying that the thickness of the sample could be on the order of microns; however, the lateral dimensions of the crystal must remain large in order to ensure a sufficient number of spins per plane for signal to noise consideration. Another requirement on the strength of the magnetic field gradient arises from the broadening of the resonances due to the recoupling of the dipolar interactions. Ideally, the dipolar broadening should be smaller than the splittings imposed by the gradient in order to prevent resonance overlap with spins on planes adjacent to the ones where dipolar interactions are selectively reintroduced. However, this need is not quite stringent; for example, with the strategy of exploiting nn couplings among nnn planes, the spins on the intermediate plane between the nnn planes exhibit no dipolar broadening due to the Lee-Goldburg broadband decoupling. In this case, overlap of resonances is avoided if the dipolar broadening is smaller than twice the splitting imposed by the static magnetic field gradient. Additional broadening due to heteronuclear dipolar interactions should also be considered. In hydroxyapatite, the $^{31}$P-$^{1}$H interactions due to the presence of the phosphate ions can be easily decoupled by selective irradiation of $^{31}$P. The $^{43}$Ca-$^{1}$H and $^{17}$O-$^{1}$H dipolar interactions can be neglected in view of the low natural abundance of $^{43}$Ca (0.145%) and $^{17}$O (0.037%).

The perspectives on this solid state NMR QIP device which we offered here can be extended in several ways in order to improve its efficiency for computational tasks. Since in a general purpose quantum computer addressing individual qubits is necessary, exploitation of parallelism by associating simultaneously independent single spin rotation, and dipolar decoupling and recoupling capabilities with each individual plane would increase the potential and flexibility of this device with only polynomial overhead. Moreover, strategies involving changes in the sign of the offset in the Lee-Goldburg irradiation scheme, and manipulation of nn couplings among different planes with $\pi$ pulses, all point to other interesting possibilities for quantum control in this device. From our point of view, our proposed design can only change for the better with advances in technology.

We are grateful for financial support for this work through grants from the Moscow Region Grant Agency Podmoscovia (Grant no. 01-01-97002) to E.B.F., and from the Natural Sciences and Engineering Research Council of Canada to S.L. We would also like to thank Dr.L.Tremblay for his help during the preparation of this manuscript.

**REFERENCES**

[1] M.A.Nielsen and I.L.Chuang, Quantum Computation and Quantum Information, Cambridge University Press, New York (2000), Chap.7.

[2] E.Knill, R.Laflamme, R.Martinez, and C.-H.Tseng, Nature **404**, 368, (2000).

[3] E.Yamaguchi and Y.Yamamoto, Appl.Phys. **A68**, 1, (1999)

[4] T.D.Ladd et al, Appl.Phys. **A71**, 27, (2000).

[5] J.R.Goldman et al, Appl.Phys.. **A71**, 11, (2000).

[6] T.D.Ladd et al, arXive e-print quant-ph/0009122, (2001).

[7] J.C.Elliott, P.E.Mackie, and R.A.Young, Science **180**, 1055, (1973).

[8] G.Cho, and J.P.Yesinowski, J.Phys.Chem. **100**, 15716, (1996), and references therein.

[9] E.B.Fel'dman and S.Lacelle, J.Chem.Phys. **107**, 7067, (1997), and references therein.

[10] S.I.Doronin et al, Chem.Phys.Lett. **341**, 597, (2001), and references therein.

[11] B.Kane, Nature **393**, 133, (1998).

[12] D.G.Cory et al, Fortschritte der Physik (in press); also available as arXive e-print quant-ph/0004104, (2000).

[13] W.Zhang and D.G.Cory, Phys.Rev.Lett. **80,** 1324, (1998).

[14] C.P.Slichter, Principles of Magnetic Resonance, Springer-Verlag, New York, 3$^{rd}$ ed., (1990).

[15] A.Blais, Phys.Rev.**A64**, 022313, (2001).

[16] J.A.Jones, R.H.Hansen, and M.Mosca, J.Magn.Reson. **135**, 353 (1998)

[17] Z.L.Mádi, R.Brűschweiler, and R.R.Ernst, J.Chem.Phys. **109**, 10603, (1998).

[18] J.L.Dodd et al, arXive e-print quant-ph/0106064, (2001).

[19] W.S.Warren, Science **277**, 1688, (1997).

[20] S.Braunstein et al, Phys.Rev.Lett. **83**, 1054, (1999).

[21] N.Linden and S.Popescu, Phys.Rev.Lett. **87**, 047901, (2001).